\documentclass[prd,preprint,superscriptaddress,tightenlines,nofootinbib, eqsecnum]{revtex4-2}

\usepackage{amsmath}
\usepackage{amsfonts}
\usepackage{amssymb}
\usepackage{bm}
\usepackage[colorlinks]{hyperref}
\usepackage{mathrsfs}
\usepackage{graphicx}
\usepackage{empheq}
\usepackage{ulem}
\usepackage{tensor}
\usepackage[usenames]{color}
\definecolor{darkgreen}{rgb}{0,0.5,0}
\hypersetup{urlcolor=darkgreen}
\usepackage[capitalize]{cleveref}

\allowdisplaybreaks

\DeclareSymbolFontAlphabet{\mathrsfs}{rsfs}
\DeclareMathAlphabet{\mathcal}{OMS}{cmsy}{m}{n}

\newcommand\calO{{\mathcal{O}}}
\newcommand{\dd}{\mathrm{d}}
\newcommand{\aem}{\alpha_\text{EM}}
\newcommand{\dJ}{\mathrm{J}}
\newcommand{\J}{\mu}

\newcommand{\go}{\mathfrak{g}}
\newcommand{\nn}{\nonumber}

\begin{document}
	
\title{Electromagnetic fields in compact binaries:\\ a post-Newtonian approach} 

\author{Quentin \textsc{Henry}}\email{quentin.henry@aei.mpg.de}
\affiliation{Max Planck Institute for Gravitational Physics\\ (Albert Einstein Institute), D-14476 Potsdam, Germany}

\author{Fran\c{c}ois \textsc{Larrouturou}}\email{francois.larrouturou@desy.de}
\affiliation{Deutsches Elektronen-Synchrotron DESY, Notkestr. 85, 22607 Hamburg, Germany}

\author{Christophe \textsc{Le Poncin-Lafitte}}\email{christophe.leponcin@obspm.fr}
\affiliation{
SYRTE, Observatoire de Paris-PSL, Sorbonne Université, CNRS UMR8630, LNE, \\ 
61 avenue de l’Observatoire, 75014 Paris, France}

\date{\today}

\preprint{DESY-23-008}

\begin{abstract}

Galactic binaries, and notably double white dwarfs systems, will be a prominent source for the future LISA and Einstein Telescope detectors.
Contrarily to the black holes observed by the current LIGO-Virgo-KAGRA network, such objects bear intense magnetic fields, that are naturally expected to leave some imprints on the gravitational wave emission.
The purpose of this work is thus to study those imprints within the post-Newtonian (PN) framework, particularly adapted to double white dwarfs systems.
To this end, we construct an effective action that takes into account the whole electromagnetic structure of a star, and then specify it to dipolar order.
With this action at hand, we compute the acceleration and Noetherian quantities for generic electric and magnetic dipoles, at a relative 2PN order.
Finally, focusing on physically relevant systems, we show that the magnetic effects on the orbital frequency, energy and angular momentum is significant, confirming previous works conclusions.

\end{abstract}

\maketitle

\section{Introduction}
\label{sec:introduction}

The first gravitational-wave event has been detected by modeling the black hole binary as two spinning, isolated points~\cite{LIGOScientific:2016aoc}.
Nevertheless, the next generations of detectors, such as the Laser Interferometer Space Antenna (LISA)~\cite{LISA:2017pwj}, or the Einstein Telescope (ET)~\cite{ET:2022}, will require finer modeling of their sources, as ten of thousands of galactic binaries, composed primarily of white dwarfs (WD) and/or neutron stars (NS), will be resolvable~\cite{Hils:1990,Timpano:2005gm}.
For such sources, taking into account environmental effects beyond the spinning point-particles approximation will be crucial. Among those effects is the presence of strong magnetic fields, that can be as intense as $10^9\, G$ for white dwarfs and $10^{12}\,G$ for neutron stars~\cite{Ferrario:2006ib,Ferrario:2020}, that may for instance modify the tidal response of the star~\cite{Zhu:2020imp}, or alter the electromagnetic signals emitted just before the merger~\cite{Wang:2018xhm}. But more importantly, those powerful magnetic fields can induce an orbital decay and shift the frequency of the gravitational wave emitted by the binary~\cite{Bourgoin:2021yzf,Bourgoin:2022ilm,Bourgoin:2022qex,Bourgoin:2022ibr,Carvalho:2022pst}.
The LISA experiment will resolve a large number of double WD~\cite{Korol:2018ulo,Korol:2021pun}, and some of those systems will even be used for the calibration of the instrument~\cite{Stroeer:2006rx,Kupfer:2018jee,Finch:2022prg}. 
It seems therefore important to incorporate the electromagnetic structure of WD and NS in the templates used by LISA and ET.

The templates that will be used for the detection and characterization of the inspiral phase of galactic binaries are mainly based on the post-Newtonian results~\cite{Mangiagli:2018kpu}, relying on both weak-field and slow-motion approximations (see \emph{e.g.}~\cite{BlanchetLR,Buonanno:2014aza,Porto:2016pyg} for reviews).
Indeed, such framework is particularly adapted to the case of those sources, as LISA will observe them deep within their inspiral phase (during which the two bodies revolve around each other, far before the merger, which happens outside the LISA frequency band).
As for ET, the galactic binaries are expected to merge within its frequency band, but after a large number of revolutions (as was the case for the binary of neutron stars observed by the LIGO-Virgo network~\cite{LIGOScientific:2017vwq}).
Therefore, post-Newtonian templates will also be extensively used for the detection and characterization of binaries of WD and/or NS in ET.

A post-Newtonian treatment of electric charges on the motion of celestial bodies widely exists in the literature, see \emph{e.g.} ~\cite{Julie:2017rpw,Khalil:2018aaj,Patil:2020dme,Gupta:2022spq,Martinez:2022vnx,Martinez:2023oga}.
But, to the best of our knowledge, a proper post-Newtonian study of magnetic effects has never been achieved, even if, as argued previously, such framework will be important for future gravitational-wave detectors.
The aim of the present work is thus to fill this gap, by building an action that consistently encompass the full electromagnetic structure of celestial bodies, and by working out explicitly the effects of dipolar electric and magnetic fields.
We will concentrate on the ``conservative'' sector (\emph{i.e.} our aim is to derive the energy, linear and angular momentum of the system), and we let the study of the radiative sector for further work.

The plan of this paper is as follows.
In Sec.~\ref{sec:action} we construct a generic action describing the interplay between gravitational interaction and electromagnetic structure for spinless bodies, and we specify it to the dipolar case.
Using this dipolar action, we derive in Sec.~\ref{sec:PN} the acceleration and Noetherian quantities, by using a post-Newtonian expansion.
Those expressions are then plugged on a quasi-circular orbit, and numerical estimates are given in Sec.~\ref{sec:quasiCirc}, before concluding in Sec.~\ref{sec:CCL}.
Lengthy expressions are displayed in App.~\ref{app:LenExpr}, and stored in an ancillary file~\cite{AncFile}.

\section{Construction of the effective action}\label{sec:action}

\subsection{General construction}\label{sec_action_gen}

The aim of this section is to construct an action, describing the interplay between the gravitational and electromagnetic (EM) interactions of two bodies (\emph{e.g.} two white dwarfs).
In order to do so, we rely on an effective description of the EM structure of each body, by ``dressing'' them up with a set of multipole moments.

In this work, and for the sake of simplicity, we will consider a point particle description of the two bodies, \emph{i.e.} neglecting the effects of spin and internal structure.
The inclusion of those effects is left for future studies.
We have to consider two dynamical fields: a metric\footnote{The conventions employed throughout this work are as follows: we use a mostly plus signature, the Minkowski metric being $\eta_{\mu\nu} = (-,+,+,+)$; greek letters denote spacetime indices $\mu,\nu,\ldots = (0,1,2,3)$ and latin ones, purely spatial indices $i,j,\ldots = (1,2,3)$; bold font denotes three-dimensional vectors, \emph{e.g.} $\boldsymbol{y}_A = y_A^i$; we use multi-index notations, \emph{i.e.} $I_L = I_{i_1i_2\ldots i_\ell}$; the d'Alembertian operator is defined with respect to the flat Minkowski metric $\Box \equiv \eta^{\mu\nu}\partial_{\mu\nu} = \Delta - c^{-2}\partial_t^2$; (anti-)symmetrizations are weighted, \emph{e.g.} $A_{(ij)} = (A_{ij} + A_{ji})/2$; the Lagrangian and Lagrangian density are denoted as $\mathcal{S} = \int\!\dd t\,\mathcal{L} =  \int\!\dd t\dd^3 x\, L$, and we will refer to ``Lagrangian'' for ``Lagrangian density'' henceforth.} $g_{\mu\nu}$ and an EM four-vector $A_\mu$
The action can thus be decomposed as
\begin{equation}
\mathcal{S} = \mathcal{S}_\text{g}[g] + \mathcal{S}_\text{EM}[A,g] + \mathcal{S}_\text{mat}\,,
\end{equation}
where $ \mathcal{S}_\text{g}$ and $ \mathcal{S}_\text{EM}$ are the kinetic terms for the metric and EM field, and $ \mathcal{S}_\text{mat}$ encodes our description of the two bodies.

\subsubsection{Gravitational action}

For the purely gravitational sector, it is natural to work with the usual Landau-Lifschitz Lagrangian, together with a gauge-fixing term (see \emph{e.g.} \cite{Bernard:2015njp})
\begin{equation}\label{eq_action_grav_pre}
L_\text{g}
= \frac{c^4}{16\pi G} \sqrt{-g}\, \bigg[g^{\mu\nu}\left(\Gamma_{\mu\rho}^\lambda\Gamma_{\nu\lambda}^\rho-\Gamma_{\mu\nu}^\lambda\Gamma_{\rho\lambda}^\rho\right)- \frac{1}{2}\,g_{\mu\nu}\Gamma^\mu\Gamma^\nu\bigg]\,,
\end{equation}
where $\Gamma^\mu_{\nu\rho}$ are the Christoffel symbols and the last term enforces the gauge $\Gamma^\mu \equiv g^{\alpha\beta}\Gamma_{\alpha\beta}^\mu = 0$.
This action differs from the usual gauge-fixed Einstein-Hilbert one only by an irrelevant total derivative, and has the advantage of containing only first derivatives of the metric.
In terms of the so-called ``gothic metric'' $\go^{\mu\nu} \equiv \sqrt{-g}\,g^{\mu\nu}$, this action becomes
\begin{equation}\label{eq_action_grav}
L_\text{g} = \frac{c^4}{32\pi G} \left[
\go_{\alpha\beta}\left(\partial_\mu \go^{\alpha\nu}\,\partial_\nu \go^{\beta\mu}-\partial_\mu\go^{\alpha\mu}\,\partial_\nu \go^{\beta\nu}\right)
- \frac{1}{2} \go^{\alpha\beta}\go_{\mu\nu}\go_{\sigma\tau}\left(
\partial_\alpha \go^{\mu\sigma}\,\partial_\beta \go^{\nu\tau}
- \frac{1}{2}\,\partial_\alpha \go^{\mu\nu}\,\partial_\beta \go^{\sigma\tau}\right)\right]\,.
\end{equation}
As usual when performing post-Newtonian (or post-Minkowskian) expansions\footnote{As usual, we dubb ``$n$PN'' a quantity of order $\calO(c^{-2n})$.}, we define the exact perturbation
\begin{equation}
h^{\mu\nu} \equiv \go^{\mu\nu} - \eta^{\mu\nu}\,.
\end{equation}
It is easy to see that the gauge relation $\Gamma^\mu = 0$ translates into the usual harmonic condition for $h^{\mu\nu}$, \emph{i.e.} $\partial_\nu h^{\mu\nu} = 0$.

\subsubsection{Electromagnetic action}

As for the EM sector, it is natural to take the gauge-fixed Maxwell Lagrangian
\begin{equation}\label{eq_action_EM_pre}
L_\text{EM} = -\frac{1}{4\mu_0}\,\sqrt{-g}\,g^{\mu\rho}g^{\nu\sigma}\bigg[F_{\mu\nu}F_{\rho\sigma} + 2 \nabla_\mu A_\rho \nabla_\nu A_\sigma\bigg]\,,
\end{equation}
where $\mu_0$ is the permeability of vacuum, and we recall that $F_{\mu\nu} \equiv 2 \nabla_{[\mu}A_{\nu]} = 2\partial_{[\mu}A_{\nu]}$ is related to the electric and magnetic fields by
\begin{equation}\label{eq_action_EB}
E_i = -c\,F_{0i}\,,\qquad\text{and}\qquad B_i  = \frac{1}{2}\, \varepsilon_{ijk}F_{jk} = \varepsilon_{ijk}\partial_jA_k\,.
\end{equation}
The last term of \cref{eq_action_EM_pre} is the gauge fixing term, imposing the Lorentz gauge on $A_\mu$, namely $\nabla^\mu A_\mu = \partial^\mu A_\mu = 0$, where we have used the harmonicity relation $\Gamma^\mu = 0$ to transform the covariant derivative into a partial one.\\

If the EM action~\eqref{eq_action_EM_pre} and the gravitational one~\eqref{eq_action_grav_pre} have the same structure, the main difference is that $\go^{\mu\nu}$ is dimensionless, when $A_\mu$ is not.
This will be a problem when performing post-Newtonian expansions, as the scaling of $h^{\mu\nu}$ and $A_\mu$ cannot  be compared.
To remedy that, we work with the dimensionless field $\tilde{A}_\mu \equiv \sqrt{e^2G/c^3\hbar}\,A_\mu$ (with $e$ the electric charge and $\hbar$ the reduced Planck constant), such that the Lagrangian becomes
\begin{equation}\label{eq_action_EM}
L_\text{EM} = -\frac{c^4}{16\pi G\,\alpha_\text{EM}}\,\sqrt{-g}\,g^{\mu\rho}g^{\nu\sigma}\bigg[\tilde{F}_{\mu\nu}\tilde{F}_{\rho\sigma}+2\nabla_\mu \tilde{A}_\rho\nabla_\nu \tilde{A}_\sigma \bigg]\,,
\end{equation}
where $\alpha_\text{EM} = \frac{\mu_0 e^2c}{4\pi\hbar}$ is the fine-structure constant, \emph{i.e.} a dimensionless number, and naturally $\tilde{F}_{\mu\nu} = 2\partial_{[\mu}\tilde{A}_{\nu]}$ .
Looking at the prefactor of~\cref{eq_action_EM}, it is clear that we can now perform consistent joint post-Newtonian expansions.
For the sake of clarity, we will henceforth drop the tildes, working only with dimensionless quantities, unless specified (when performing the matching in Sec.~\ref{sec_action_matching}, and naturally, when performing numerical estimates in Sec.~\ref{subsec:AN}).

\subsubsection{Matter action}

As advertised, we consider two spinless stars, having some electromagnetic structure.
In order to model those structures, we rely on a effective approach: as seen by a distant observer, those stars can be described by point particles,``dressed'' with  a set of EM moments.

Without any EM structure, the action of a point-particle $A$ is simply given by its proper time $\tau_A$, the coupling constant being its mass $m_A$:
\begin{equation}\label{eq_action_pp}
\mathcal{S}_\text{pp} = -c\,\sum_A\,m_A \int\!\!\dd\tau_A\,.
\end{equation}
As for the EM sector, it is described by the interaction between the EM field and a conserved current $j^\mu_A$
\begin{equation}\label{eq_action_jA}
\mathcal{S}_\text{EM,inter} =\sum_A \int\!\!\dd^4x\,\sqrt{-g}\, j^\mu_A A_\mu\,.
\end{equation}
The current $j^\mu_A$ encompasses the whole physics of the EM structure of the star, and thus its knowledge requires stellar physics that is largely beyond the scope of this work.
Therefore, we seek for an effective description of this interaction.
To this purpose, we assume that the source $j^\mu_A$ is compact supported and that we are in the long wavelength approximation, \emph{i.e.} that the typical scale of variation of $A_\mu$ is much larger than the size of the source (as will be the case when dealing with a binary of stars).
In such configuration, we can perform a derivative expansion around a point within the source, following the derivation of~\cite{Ross:2012fc}.
We thus expand the action~\eqref{eq_action_jA} as
\begin{equation}\label{eq_action_matter_EM}
\mathcal{S}_\text{EM,eff} = 
\sum_A\,\int\!\!\dd\tau_A
\,\bigg[q_A\, A_0+ \sum_{\ell\geq 1}\frac{(-)^\ell}{\ell !}\left(\frac{I_A^L}{c}\,\nabla^\perp_{L-1}E_{i_\ell}+ \frac{\ell\,J_A^L}{\ell+1}\,\nabla^\perp_{L-1}B_{i_\ell}\right)\bigg]\,,
\end{equation} 
where $\nabla^\perp$ is the covariant derivative projected onto a set of spatial vielbein, $(E_i,B_i)$ are the electric and magnetic fields defined in \cref{eq_action_EB}.
The information contained in $j^\mu_A$ is now dispatched onto two infinite sets of time-dependent moments: electric ones $\{I_A^L\}$ and magnetic ones $\{J_A^L\}$, as well as a conserved charge $q_A$, corresponding to the electric monopole.
As usual when dealing with effective actions, the moments $(I_A^L,J_A^L)$ are not determined here, but require some matching to reveal the physical information they bear.
The main advantage of such an action is that we can focus on the physical effect we want to consider.
As we are interested by the effect of the EM dipolar structure of the stars, we will hereafter select only the $\ell=1$ terms.\\

Adding \cref{eq_action_grav,eq_action_EM,eq_action_pp,eq_action_matter_EM}, we obtain a consistent action describing the  the interplay between gravitational and EM interactions of two stars, thus fulfilling the goal of this section.

\subsection{Matching of the dipolar order}\label{sec_action_matching}

Having properly defined an action to work with, let us now focus on the aim of this work: the dipolar order $\ell = 1$.
As clear from \cref{eq_action_matter_EM}, the dipoles couple to single derivatives of $A_\mu$, thus we can collect them into an anti-symmetric four-tensor $\mathcal{D}^{\mu\nu}$ that enters the action through a coupling $\propto F_{\mu\nu}\mathcal{D}^{\mu\nu}$.
As we seek to extract physical information from our computation, we need to link the two sets of dipoles $\{I_A^i,J_A^i\}$ (or equivalently $\mathcal{D}^{\mu\nu}$) to the observable electric and magnetic moments of the stars $\{q_A^i,\mu_A^i\}$.
To do this matching, we work with dimensionfull quantities, and set the gravitational interaction and the masses to zero.
We thus consider the Lagrangian
\begin{equation}
L_\text{match} =- \frac{1}{4\mu_0}\bigg[F_{\mu\nu}F^{\mu\nu}+2 \left(\partial_\mu A^\mu\right)^2\bigg]+\sum_A\,\frac{\delta_A}{2}\,\left(F_{\mu\nu}\right)_A \mathcal{D}_A^{\mu\nu}\,,
\end{equation}
where $\delta_A\equiv \delta[x^i-y_A^i(t)]$ is the 3-dimensional Dirac delta distribution, locating the interaction on the worldline of the particle, $\boldsymbol{y}_A(t)$, and the subscript $A$ means that the quantity is regularized in the worldline of particle $A$.
Performing a 3+1 decomposition (which is trivial since we have no gravitational interaction), the equations of motion for the field $A_\mu$ read
\begin{equation}
\Box A_0 = -\mu_0\,\mathcal{D}_{0k}\,\partial_k\delta_A\,,
\qquad\text{and}\qquad
\Box A_i = \frac{\mu_0}{c}\,\frac{\dd}{\dd t}\bigg[\mathcal{D}_{0i}\,\delta_A\bigg] - \mu_0\,\mathcal{D}_{ik}\,\partial_k\delta_A\,.
\end{equation}
Integrating them, and denoting $r = \vert \boldsymbol{y}_1(t)- \boldsymbol{y}_2(t)\vert$ and $n^i = [y_1^i(t)-y_2^i(t)]/r$, it comes
\begin{equation}
A_0 = - \frac{\mu_0}{4\pi\,r^2}\,n^k\mathcal{D}_{0k}\,,
\qquad\text{and}\qquad
A_i =  - \frac{\mu_0}{4\pi\,r^2}\,n^k\mathcal{D}_{ik}  - \frac{\mu_0}{4\pi\,c}\frac{\dd}{\dd t}\bigg[\frac{\mathcal{D}_{0i}}{r}\bigg]\,.
\end{equation}
Comparing this result to the usual formula for constant dipoles~\cite{jackson1999classical}
\begin{equation}
A_0 = - \frac{\mu_0\,c\,q^k}{4\pi}\,\frac{n^k}{r^2}\,,
\qquad\text{and}\qquad
A_i =  - \frac{\mu_0\,\dJ^{ik}}{4\pi}\,\frac{n^k}{r^2}\,,
\end{equation}
allows to fix
\begin{equation}\label{eq_action_Dmunu}
\mathcal{D}^{\mu\nu}_A =
\begin{pmatrix}
0 & -c\,q^i_A\\
c\,q^i_A & \dJ^{ij}_A
\end{pmatrix}\,,
\end{equation}
where we have shortened $\dJ_A^{ij} \equiv \varepsilon_{ijk}\mu_A^k$.
In the following, we will thus use the matter action
\begin{equation}\label{eq_action_mat}
\begin{aligned}
\mathcal{S}_\text{mat} 
& =
-\sum_A\int\!\!\dd\tau_A
\,\bigg[
m_A c^2- \frac{1}{2}\,\left(F_{\mu\nu}\right)_A \mathcal{D}^{\mu\nu}_A\bigg]\\
& = 
-\sum_A\int\!\!\dd t\,\dd^3x\,\frac{\delta_A}{u_A^0}
\,\bigg[
m_A c^2
-c\,q_A^k\left(\partial_kA_0\right)_A+q_A^k\left(\partial_tA_k\right)_A-\dJ^{ij}_A\,\left(\partial_iA_j\right)_A\bigg]\,,
\end{aligned}
\end{equation} 
where $u_A^0 = [-(g_{\mu\nu})_A\,v_A^\mu v_A^\nu/c^2]^{-1/2}$ is the Lorentz factor, $v_A^\mu = (c,v_A^i)$ (with $v_A^i$ the usual 3-dimensional velocity). To obtain the second line of~\eqref{eq_action_mat}, we performed a 3+1 decomposition by using the relation
\begin{equation}
\dd\tau_A = \dd t \int\!\dd^3x\, \frac{\delta_A}{u_A^0}\,.
\end{equation}
In~\cref{eq_action_mat}, the moments have naturally been renormalised by $(q_A^i,\dJ_A^{ij}) \to \sqrt{\frac{c^3\hbar}{G\,e^2}}\,(q_A^i,\dJ_A^{ij})$, following the normalisation of $A_\mu$.\\

Another, fully equivalent, way to obtain the matter action is to consider that linearity and parity allow us to write two and only two interactions: $c_E\,q_A^i E_i$ and $c_B\,\mu_A^i B_i$, where $c_E$ and $c_B$ are yet unknown coefficients.
Solving for $A_\mu$ and matching the result to the usual values allows to fix them and, writing the electric and magnetic fields in terms of $A_\mu$, one recovers~\cref{eq_action_mat}.

\section{Post-Newtonian computation of the electromagnetic dipolar effects}\label{sec:PN}

In a perturbative problem, one can use the Fokker method to build a Lagrangian equivalent to the original one.
The idea of this method is to vary the Lagrangian with respect to the fields first (in our case, $h^{\mu\nu}$ and $A_\mu$), solve the field equations in terms of the matter degrees of freedom (position velocities, \emph{etc}), and then inject the solutions into the original Lagrangian.
The resulting (Fokker) Lagrangian only depends on the matter degrees of freedom, and solving the associated equations of motion is totally equivalent to solving for the initial Lagrangian.
This procedure is analogue to the method of integrating out the fields, commonly used in the effective field theory framework.

We are interested in computing the next-to-leading order (NLO) of the magnetic dipole effects, which appear at $\calO(c^{-4})$ in the action (as proven below).
However, the electric dipole arises at $\calO(c^{-2})$ in the action, so we need to take into account the next-to-next-to-leading order (NNLO) electric contributions to be consistent for the NLO magnetic effects.
This means that we have to push the usual point-particle sector up to 2PN to be coherent with this NNLO computation.\footnote{All the computations presented in this section and in the next one were performed with the use of the \textit{xAct} library from the \textit{Mathematica} software~\cite{xtensor}.}

\subsection{Field equations}\label{subsec:field}

Let us start the Fokker procedure, and derive the field equations of the problem. To do so, we vary the total Lagrangian, composed of~\cref{eq_action_grav,eq_action_EM,eq_action_mat} with respect to both the metric $g_{\mu\nu}$ (or equivalently, the gothic metric $\go^{\mu\nu}$) and the EM field $A_\mu$. By imposing the cancellation of the variations with respect to the two fields $\delta L /\delta g_{\mu\nu} = 0 = \delta L /\delta A_\mu$, one finds the following wave equations verified by the perturbed metric and the EM field
\begin{subequations}\label{eq:fields}
\begin{align}
\Box h^{\mu\nu} &= \mathcal{T}^{\mu\nu} + \Lambda^{\mu\nu}\,,\\
\Box A_\mu &= \mathcal{V}_\mu + \Phi_\mu\,.
\end{align}
\end{subequations}
We have split the source of the wave equations into a compact part ($\mathcal{T}^{\mu\nu},\mathcal{V}_\mu$, located on the worldline of the particles) and non-compact part ($\Lambda^{\mu\nu}, \Phi_\mu$, extending in the whole space). The compact contributions of the sources of the d'Alembert equations read
\begin{subequations}
\begin{align}
\mathcal{T}^{\mu\nu} &= \frac{16\pi G}{c^4}\sum_A \sqrt{\vert g\vert_A}\, u_A^0 v_A^\mu v_A^\nu \left(m_A-\frac{1}{2 c^2}\left(F_{\mu\nu}\right)_A \mathcal{D}^{\mu\nu}_A\right)\delta_A\,,\\
\mathcal{V}_\mu &= -\frac{4\pi G \aem}{c^4}g_{\mu\nu}\sum_A \partial_\lambda\left( \frac{\mathcal{D}^{\nu\lambda}_A}{u_A^0}\delta_A \right)\,. 
\label{eq_PN_vmu_expr}
\end{align}
\end{subequations}
We recall that the subscript $A$ indicates that the quantities are regularized on the location of body $A$, which mean that they are functions of the time parameter $t$ only. Regarding the non-compact quantities, one can split $\Lambda^{\mu\nu} = \Lambda^{\mu\nu}_\text{g} + \Lambda^{\mu\nu}_\text{EM}$ into a sum of the purely gravitational contribution $\Lambda^{\mu\nu}_\text{g}$ and terms containing EM effects $\Lambda^{\mu\nu}_\text{EM}$. The expression of $\Lambda^{\mu\nu}_\text{g}$ can be found in, e.g. Eq.~(24) of \cite{BlanchetLR}. The rest reads
\begin{subequations}
\begin{align}
\Lambda^{\mu\nu}_\text{EM} &= \frac{4|g|}{\aem}\left(F^{\mu\lambda}F\indices{^\nu_\lambda}-\frac{g^{\mu\nu}}{4}F_{\alpha\beta}F^{\alpha\beta} \right) \,,\\
\Phi_\mu &= - h^{\alpha\beta}\partial_{\alpha\beta}A_\mu  - \partial_\mu \go^{\alpha\beta} \,\partial_\alpha A_\beta - \go_{\mu\nu} \go^{\alpha\lambda}\,F_{\alpha\beta}\partial_\lambda\go^{\beta\nu} + \frac{1}{2}\,\go^{\alpha\beta}\go_{\lambda\rho}\,F_{\alpha\mu}\partial_\beta\go^{\lambda\rho}\,.
\end{align}
\end{subequations}
We recover for $\Lambda^{\mu\nu}_\text{EM}$ the usual expression of the EM stress-energy tensor. Note that these expressions are exact in the sense that we did not perform any truncation in powers of the perturbed metric. However, we extensively used the gauge conditions $\partial_\nu h^{\mu\nu} = 0$ and $\nabla^\mu A_\mu =\partial^\mu A_\mu = 0$ to derive them.\\

In order to solve Eqs.~\eqref{eq:fields} within the PN approach, we need to perform a 3+1 decomposition, \emph{i.e.} differentiate spatial from temporal indices so that the factors in $c$ are explicit. Thus, we express Eqs.~\eqref{eq:fields} in terms of the perturbed metric $h^{\mu\nu}$, then truncate the expressions to the cubic order in $h^{\mu\nu}$. Finally, we separate spatial and temporal indices.
Plugging the expression of $\mathcal{D}^{\mu\nu}$~\eqref{eq_action_Dmunu} into the compact source for $A_\mu$~\eqref{eq_PN_vmu_expr}, it comes that $\mathcal{V}_0$ starts at $\calO(c^{-3})$ and $\mathcal{V}_i$, at $\calO(c^{-4})$.
As the d'Alembertian operator does not change the leading PN order, $A_\mu$ will also start at $\calO(c^{-3},c^{-4})$.
Injecting those orders back into the action, it is clear that the electric effects starts at $\calO(c^{-2})$ in the action, and the magnetic ones, at $\calO(c^{-4})$ (see \emph{e.g.}~\cref{eq_action_mat}).
Therefore, we truncate the expressions to $\calO(c^{-4})$ for the point-particle part and to $\calO(c^{-6})$ for the EM part, which represents the NNLO for both point-particle and EM effects.

\subsection{Solving the field equations}\label{subsec:metric}
\subsubsection{Metric and EM field parametrizations}\label{subsubsec:param}

Once the field equations~\eqref{eq:fields} derived, we have to solve them in terms of the matter variables.
To simplify the procedure, we expand the metric and EM field in PN orders, and parametrize each order with some PN potentials.
Solving for the fields thus turns out to be equivalent to solving for those elementary potentials.
In order to be able to efficiently check our computation against the knwon point-particle derivation, we have chosen our parametrization in such a way that if one sets EM effects to zero, we recover the usual PN metric at the 2PN order~\cite{Blanchet:2000ub}. The parametrized metric  reads
\begin{subequations}\label{eq:PNmetric}
\begin{align}
h^{00} &= -\frac{4V}{c^2} - \frac{4}{c^4}\left[2V^2+\frac{W_{kk}}{2}\right]- \frac{4}{c^6}\left[4X+2Z_{kk}+\frac{8V^3}{3}+2V W_{kk}- \frac{\varphi^2}{2}\right] \\
&\qquad+ \frac{6}{c^8}\,\varphi^2V+ \mathcal{O}(c^{-10})\nn\,,\\
h^{0i} &= - \frac{4V_i}{c^3}- \frac{8}{c^5}\bigg[R_i+V V_i\bigg]+\mathcal{O}(c^{-7})\,,\\
h^{ij} &= -\frac{4}{c^4}\left(W_{ij}-\frac{\delta_{ij}}{2}\, W_{kk}\right)-\frac{16}{c^6}\left(Z_{ij}-\frac{\delta_{ij}}{2}\, Z_{kk}\right)+ \mathcal{O}(c^{-8})\,.
\end{align}
\end{subequations}
In this parametrization, we find the usual PN potentials $\{V,V_i,W_{ij},X,R_i,Z_{ij}\}$ to which we added a new, purely EM, potential $\varphi$. We then chose the parametrization of the EM field as
\begin{subequations}\label{eq:PNem}
\begin{align}
A_0 &= \frac{\varphi}{c^3}-\frac{V\varphi}{c^5}+ \frac{1}{c^7}\bigg[\sigma+V^2\varphi\bigg] +\mathcal{O}(c^{-9})\,,\\
A_i &= \frac{\chi_i}{c^4} + \frac{1}{c^6}\bigg[\psi_i+4\varphi V_i\bigg] +\mathcal{O}(c^{-8})\,,
\end{align}
\end{subequations}
where we introduced the additional EM potentials $\{\chi_i,\sigma,\psi_i\}$. The gauge conditions on the metric and EM fields $\partial_\nu h^{\mu\nu} = 0$ and $\partial^\mu A_\mu = 0$ induce some relations between the potentials.
At the required order, it comes
\begin{subequations}\label{eq:EMharmpots}
\begin{align}
& \partial_t V + \partial_i V_i = \calO(c^{-2})\,,\\
& \partial_tV_i + \partial_kW_{ik} - \frac{1}{2}\,\partial_i W_{kk} = \calO(c^{-2})\,,\\
& \partial_t \varphi- \partial_i\chi_i + \frac{1}{c^2}\bigg[3V\partial_t\varphi +3 \varphi\,\partial_t V - \partial_i\psi_i\bigg] = \calO(c^{-4})\,.
\end{align}
\end{subequations}

\subsubsection{Computation of the potentials}\label{subsubsec:pots}

Inserting the parametrized metric~\eqref{eq:PNmetric} and EM field~\eqref{eq:PNem} into the field equations~\eqref{eq:fields}, we obtain a set of wave equations, to be satisfied by the potentials. Naturally, the sources for the usual ``point-particle'' potentials include now EM contributions. Moreover, regarding the purely EM potentials $\{\varphi,\chi_i,\sigma,\psi_i\}$, we exploit the freedom on the definitions of the sources to include all compact terms in the leading order potentials $\varphi$ and $\chi_i$, and the non-compact terms into the sources of $\psi_i$ and $\sigma$.
The wave equations read
\begin{subequations}\label{eq:potdefs}
\begin{align}
&\Box V = -4\pi G \sum_A \Bigl[\xi_A^i\delta_A + \overline{\xi}_A^i\partial_i\delta_A \Bigr]\,,\\
&\Box V_i = -4\pi G \sum_A \zeta_A v_A^i \delta_A -\frac{\big(\partial_i\chi_j-\partial_j\chi_i\big)\partial_j\varphi}{\aem c^4}\,,\\
&\Box W_{ij} = -4\pi G \sum_A\lambda_A\big(v_A^{ij}-v_A^2\delta^{ij}\big)\,\delta_A-\partial_iV\partial_j V+\frac{\partial_i\varphi\partial_j\varphi}{\aem c^2}\,,\\
&\Box \varphi = -4\pi G \aem\sum_A\big(\omega_A \delta_A+\overline{\omega}_A^i\partial_i\delta_A \big)\,,\\
&\Box \chi_i = -4\pi G\,\aem \sum_A\big(\theta_A^i \delta_A+\overline{\theta}_A^{ij}\partial_j\delta_A \big)\,,\\
&\Box \sigma = 4W_{ij}\partial_{ij} \varphi - 4\big(\partial_i \chi_j -\partial_j \chi_i \big)\partial_i V_j +2\partial_t\chi_i\partial_i V +8V_i\partial_t\partial_i\varphi\\
& \qquad\quad\nonumber
-2V\partial_iV\partial_i\varphi +2\partial_t V\partial_t\varphi+4V\partial_t^2\varphi \,,\\
&\Box \psi_i = 4 \big(\partial_iV_j-\partial_jV_i\big)\partial_j\varphi-2 \big(\partial_i\chi_j-\partial_j\chi_i\big)\partial_jV+4\partial_t\varphi\partial_i V+2\partial_tV\partial_i\varphi\,,
\end{align}
\end{subequations}
where the quantities entering the definitions of the sources are given in Eq.~\eqref{eq:potsourcesdef}.
The sources of the potentials $X$, $R_i$ and $Z_{ij}$ entering the metric~\eqref{eq:PNmetric} are not displayed here, as those potentials will play no role hereafter, \emph{c.f.} next section (nevertheless, the curious reader can refer \emph{e.g.} to App.~A of~\cite{Marchand:2020fpt} for their point-particle expressions).
The reason why we chose to put all complicated, non-compact terms into the sources of $\psi_i$ and $\sigma$ (instead of including them in $\varphi$ and $\chi_i$) is because, as we expect from the Fokker method and will see in Sec.~\ref{subsec:Fokker}, we do not need to compute all the potentials in order to derive the Fokker Lagrangian. Notably those two potentials will play no role, as clear from Table~\ref{table_pots}, showing the potentials and their order required to control the Fokker Lagrangian to NNLO.

\begin{table}[h]
\centering
\begin{tabular}{|c||c|c|c|c|c|c|c|c|c|c|c|}
\hline
& $V$ & $V_i$ & $W_{ij}$ & $\varphi$ & $\sigma$ & $\chi_i$ & $\psi_i$ \\
 \hline
 \hline
0PN  & $\times$ & $\times$ & $\times$ & $\times$ & & $\times$ & \\
 \hline
1PN & $\times$ & & & $\times$ & & $\times$ & \\
 \hline
2PN & & & & $\times$ & & & \\
 \hline
\end{tabular}
\caption{Required potentials and their absolute PN order needed to compute the Fokker Lagrangian. The potentials $X$, $R_i$ and $Z_{ij}$ entering the metric~\eqref{eq:PNmetric} play no role in the following computation, and thus are not displayed here. Note in particular that we do not need to derive the EM part of the potentials $V$, $V_i$ and $W_{ij}$, as they enter at higher than needed PN orders.}
\label{table_pots}
\end{table}

The method used to integrate the waves equations~\eqref{eq:potdefs} to the required PN order with \textit{Hadamard partie-finie} regularization are well-known, and described in, \emph{e.g.}~\cite{Blanchet:1998vx}. 
The lengthy expressions of the potentials are not very enlightening \emph{per se} and thus will not be displayed here. 
Note however that we have explicitly checked that the gauge conditions~\eqref{eq:EMharmpots} are satisfied by the computed potentials.

\subsection{Fokker Lagrangian and equations of motion}\label{subsec:Fokker}

As mentioned above, the idea of the Fokker method is to insert the explicit expression of the fields back into the action. This defines a new Lagrangian, equivalent to the original one. To do so, we start from the 3+1 decomposed and PN expanded Lagrangian resulting from Sec.~\ref{subsec:field} and insert the metric~\eqref{eq:PNmetric} and EM field~\eqref{eq:PNem}, without (yet) replacing the potentials by their values. This gives a Lagrangian, written in terms of potentials, which we simplify by performing integrations by parts, making use of the gauge relations and replacing d'Alembertian operators acting on potentials by the expression of the corresponding sources. Those manipulations result is the disappearance of the potentials $X$, $R_i$, $Z_{ij}$, $\psi_i$ and $\sigma$, as well as the EM parts of the potentials $V$, $V_i$ and $W_{ij}$. In the end, the only potentials required are displayed in Table~\ref{table_pots}.\\

One can then replace the remaining potentials by their explicit expressions, and the only step remaining is to perform  integrals of the form $\int\! \dd^3 \mathbf{x}\, S(\mathbf{x},t)$. 
Two types of integrals appear: compact ones, proportional to (derivatives) of Dirac distributions, and non-compact ones, extending over all space.
The compact sector of the Lagrangian is computed by regularizing the integrand using the \textit{Hadamard partie-finie} regularization~\cite{Blanchet:2000nu}. On the other hand, the non-compact part of the Lagrangian is treated with the method described \emph{e.g.} in~\cite{Blanchet:2003gy}. However, the computations have to be performed in the sense of distributions, and the non-compact terms also induce compact supported integrals, through distributional derivatives. The most simple example is $\partial_{ab}(1/r_1) = 3 \hat{n}_1^{ab}r_1^{-3} - \tfrac{4\pi}{3}\delta_{ab}\delta_1$ (see Sec.~IV C of~\cite{Marchand:2020fpt} and references therein for a discussion on the computation of those terms). While this does not appear in the purely point-particle case, it is crucial to take them into account for the EM part. For instance, the distributional part of the spatial derivative of $\varphi$ reads at leading order $(\partial_i \varphi)_\text{distr} = -\tfrac{4\pi}{3}\aem G q_1^i\delta_1 + (1 \leftrightarrow 2)$. As those distributional derivatives are proportional to (derivatives) of Dirac distributions, they are computed as the regular compact-supported integrals.\\

After computing all the integrals constituting the Fokker Lagrangian, we obtain a generalized Lagrangian depending on the dynamical variables of the system. This Lagrangian depends on positions, velocities and accelerations, as well as derivatives of accelerations. We can now apply the generalized Euler-Lagrange equation which allows us to derive the accelerations of bodies 1 and 2. Once the equations of motion are known, we apply some reduction methods, described in \emph{e.g.}~\cite{deAndrade:2000gf} to obtain a reduced Fokker Lagrangian $\mathcal{L}$ that only depends on positions, velocities, accelerations, and that is linear in accelerations.
As for the EM moments, the reduced Fokker Lagrangian depends on (at most) second derivatives of $q_A^i$, and (at most) first derivatives of $\dJ_A^{ij}$.
Both generalized and reduced Fokker Lagrangian are naturally equivalent, in the sense that they give the same acceleration and Noetherian quantities.

\subsection{Invariance of the action and Noetherian quantities}\label{subsec:Noether}

Once the Fokker Lagrangian $\mathcal{L}$ has been computed, it is important to verify that it is a scalar under the Poincar{\'e} group.
Let us recall the way it has to behave under the ten Poincar{\'e} transformation
\begin{subequations}\label{eq_PN_Noether_transfo_L}
\begin{align}
&
\text{spatial translation} & &
x^i \to x^i + \epsilon^i\qquad
& & \Rightarrow \quad \delta_\epsilon \mathcal{L} = 0\,, \\
&
\text{spatial rotation} & &
x^i \to  R^i_{\ j}(\omega)\,x^j\qquad
& & \Rightarrow  \quad\delta_\omega \mathcal{L} = 0\,,\\
&
\text{temporal translation} & &
t \to t + \tau \qquad
& & \Rightarrow \quad \delta_\tau \mathcal{L} = \frac{\dd \mathcal{L}}{\dd t}\,\tau\,,\\
&
\text{Lorentz boost} & &
x^\mu \to \Lambda^\mu_{\ \nu}(\beta)\, x^\nu \qquad
& & \Rightarrow  \quad\delta_\beta \mathcal{L} = \frac{\dd Z^i}{\dd t}\,\beta^i\,,
\end{align}
\end{subequations}
where $R^i_{\ j}(\omega)$ is a rotation matrix of parameter $\omega^i$, $\Lambda^\mu_{\ \nu}(\beta)$ is a boost matrix of parameter $\beta^i$, and $Z^i$ is an unconstrained quantity. Using the fact that $\mathcal{D}^{\mu\nu}$, defined in~\cref{eq_action_Dmunu}, is a Lorentz tensor (as $F_{\mu\nu} \mathcal{D}^{\mu\nu}$ is a Lorentz scalar), and thus transforms under a boost as
\begin{equation}
\mathcal{D}^{\mu\nu} \to \Lambda^\mu_{\ \rho}\Lambda^\nu_{\ \sigma}\,\mathcal{D}^{\rho\sigma}
\qquad\text{where}\qquad
\Lambda^\mu_{\ \nu} =
\begin{pmatrix}
1 & - \frac{\beta^i}{c}\\
- \frac{\beta^i}{c} & \delta_{ij}
\end{pmatrix}
+
\mathcal{O}\left(\beta^2\right)\,,
\end{equation}
the quantities entering the Lagrangian behave under infinitesimal transformations as
\begin{subequations}\label{eq_PN_Noether_transfo_qtt}
\begin{align}
&
\delta_\epsilon y_A^i = \epsilon^i\,, & &
\delta_\epsilon q_A^i = 0\,, & & 
\delta_\epsilon \dJ_A^{ij} = 0\,, \\
&
\delta_\omega y_A^i = \varepsilon^{ijk}y_A^j\omega^k\,, & &
\delta_\omega q_A^i = \varepsilon^{ijk}q_A^j\omega^k\,, & &
\delta_\omega \dJ_A^{ij} = \big(\varepsilon^{iak}\dJ_A^{aj}-\varepsilon^{jak}\dJ_A^{ai}\big)\omega^k\,,\\
&
\delta_\tau y_A^i = v_A^i\,\tau\,, & &
\delta_\tau q_A^i = \dot{q}_A^i\,\tau\,, & & 
\delta_\tau \dJ_A^{ij} = \dot{\dJ}_A^{ij}\,\tau\,,\\
&
\delta_\beta y_A^i = - \beta^i t+ \frac{\beta^ky_A^k}{c^2}v_A^i\,, & &
\delta_\beta q_A^i = -\dJ_A^{ij}\,\frac{\beta^j}{c^2}+\frac{\beta^ky_A^k}{c^2}\,\dot{q}_A^i\,, & &
\delta_\beta\dJ_A^{ij} = \beta^iq_A^j-\beta^jq_A^i
+ \frac{\beta^ky_A^k}{c^2}\,\dot{\dJ}_A^{ij}\,,
\end{align}
\end{subequations}
where the variations under the boost have been evaluated at the original time $t$~\cite{Blanchet:2000cw}.
Injecting the variations~\eqref{eq_PN_Noether_transfo_qtt} into the expression of the Lagrangian, one recovers indeed the expected behavior~\eqref{eq_PN_Noether_transfo_L}, with $Z^i$ displayed in~\cref{eq_lengthy_Zi}.

Now that we have checked that the Lagrangian behaves nicely under the Poincar{\'e} group, one can extract the ten Noetherian quantities, corresponding to the ten transformations: the Hamiltonian $\mathcal{H}$, associated with the temporal translation, the linear and angular momenta $\mathcal{P}^i$ and $\mathcal{J}^i$, associated respectively with the spatial translation and rotation, and the center-of-mass integral $\mathcal{G}^i$, associated with the Lorentz boost.
Following~\cite{deAndrade:2000gf}, we define the generalized moments\footnote{Recall that our Lagrangian depends on positions and velocities, but also on accelerations.}
\begin{equation}
\pi_A^i \equiv \frac{\partial \mathcal{L}}{\partial v_A^i}-\frac{\dd}{\dd t}\bigg[\frac{\partial \mathcal{L}}{\partial a_A^i}\bigg]\,,\qquad\text{and}\qquad
\theta_A^i \equiv \frac{\partial \mathcal{L}}{\partial a_A^i}\,,
\end{equation}
where the acceleration is to be replaced by its \emph{on-shell} value.
The Noetherian quantities are then defined as~\cite{deAndrade:2000gf}
\begin{subequations} \label{eq_PN_Noether_def}
\begin{align}
&
\mathcal{H} \equiv \sum_A\big[\pi_A^kv_A^k+\theta_A^ka_A^k\big]-\mathcal{L}\,,\label{eq_PN_Noether_def_H}\\
&
\mathcal{P}^i \equiv \sum_A\pi_A^i\,, \label{eq_PN_Noether_def_Pi}\\
&
\mathcal{J}^i \equiv \varepsilon_{ijk} \sum_A\big[y_A^j\pi_A^k+v_A^j\theta_A^k\big]\,,\label{eq_PN_Noether_def_Ji}\\
&
\mathcal{G}^i \equiv -Z^i -  \sum_A\bigg[\theta_A^i - \frac{1}{c^2}\big(y_A^i\pi_A^kv_A^k+y_A^i\theta_A^ka_A^k+v_A^i\theta_A^kv_A^k\big)\bigg]\,.\label{eq_PN_Noether_def_Gi}
\end{align}
\end{subequations}
Note here an important point: we dubbed those quantities as ``Noetherian'' and not as ``conserved''.
Indeed, due to the presence of the non-dynamical moments $\{q_A^i,\dJ_A^{ij}\}$, they are not conserved.
If the dynamics of the EM moments were to be fixed (for instance, by setting the magnetic moment to be proportional to the spin of the star, as done in~\cite{Bourgoin:2022ilm}), the definitions of the Noetherian quantities~\eqref{eq_PN_Noether_def} would bear extra pieces, corresponding to this additional dynamics.
In this case, those newly defined quantities would be conserved.
However, in our case, the EM moments are totally free, but obey non-trivial transformations under the Poincar{\'e} group (except under spatial translations), as clear from~\cref{eq_PN_Noether_transfo_qtt}.
Therefore (excepted for the linear momentum $\mathcal{P}^i$), the Noetherian quantities are not conserved but rather obey \emph{flux-balance equations}, namely
\begin{subequations}\label{eq_PN_fluxbalance}
\begin{align}
&
\frac{\dd \mathcal{H}}{\dd t} \equiv - \mathcal{F}\,,
& \qquad & 
\frac{\dd \mathcal{P}^i}{\dd t} \equiv 0\,,\\
&
\frac{\dd \mathcal{J}^i}{\dd t} \equiv - \Upsilon^i\,,
& \qquad & 
\frac{\dd \mathcal{G}^i}{\dd t} \equiv \mathcal{P}^i - \Psi^i\,,
\end{align}
\end{subequations}
where the fluxes are given by (we recall that $\mathcal{L}$ contains at most second derivatives of $q_A^i$, and first derivatives of $\dJ_A^{ij}$)
\begin{subequations}\label{eq_PN_flux_expr}
\begin{align}
\mathcal{F} & \label{eq_PN_flux_expr_F}
= \frac{\partial \mathcal{L}}{\partial t}
= \sum_A\bigg[
\dot{q}_A^i\frac{\partial \mathcal{L}}{\partial q_A^i}
+\ddot{q}_A^i\frac{\partial \mathcal{L}}{\partial \dot{q}_A^i}
+\dddot{q}_A^i\frac{\partial \mathcal{L}}{\partial \ddot{q}_A^i}
+\dot{\dJ}_A^{ij}\frac{\partial \mathcal{L}}{\partial \dJ_A^{ij}}
+\ddot{\dJ}_A^{ij}\frac{\partial \mathcal{L}}{\partial\dot{\dJ}_A^{ij}}\bigg]\,,\\
\Upsilon^i & \label{eq_PN_flux_expr_Upsilon}
= - \delta_\omega^\text{EM}\mathcal{L}
= \varepsilon_{ijk} \sum_A\bigg[
q_A^j\frac{\partial \mathcal{L}}{\partial q_A^k}
+\dot{q}_A^j\frac{\partial \mathcal{L}}{\partial \dot{q}_A^k}
+\ddot{q}_A^j\frac{\partial \mathcal{L}}{\partial \ddot{q}_A^k}
+2\,\dJ_A^{ja}\frac{\partial \mathcal{L}}{\partial \dJ_A^{ka}}
+2\,\dot{\dJ}_A^{ja}\frac{\partial \mathcal{L}}{\partial \dot{\dJ}_A^{ka}}\bigg]\,,\\
\Psi^i & \nonumber
= \delta_\beta^\text{EM}\mathcal{L}
= 
\frac{1}{c^2}\sum_A\bigg[\frac{\partial \mathcal{L}}{\partial q_A^k}\big(\dot{q}_A^ky_A^i-\dJ_A^{ki}\big)
+\frac{\partial \mathcal{L}}{\partial \dot{q}_A^k}\frac{\dd\big(\dot{q}_A^ky_A^i-\dJ_A^{ki}\big)}{\dd t}
+\frac{\partial \mathcal{L}}{\partial \ddot{q}_A^k}\frac{\dd^2\big(\dot{q}_A^ky_A^i-\dJ_A^{ki}\big)}{\dd t^2}\bigg]\\
& \qquad\qquad\quad
+ \sum_A \bigg[\frac{\partial \mathcal{L}}{\partial \dJ_A^{ab}}\bigg(\delta^{ia}q_A^b-\delta^{ib}q_A^a +\frac{\dot{\dJ}_A^{ab}y_A^i}{c^2}\bigg)
+ \frac{\partial \mathcal{L}}{\partial \dot{\dJ}_A^{ab}}\bigg(\delta^{ia}\dot{q}_A^b-\delta^{ib}\dot{q}_A^a +\frac{\ddot{\dJ}_A^{ab}y_A^i+\dot{\dJ}_A^{ab}v_A^i}{c^2}\bigg)\bigg]\,.
\end{align}
\end{subequations}
Note that if we take the EM moments to be constant in time, then the Hamiltonian flux $\mathcal{F}$ vanishes, as expected.
Naturally, the flux-balance equations~\eqref{eq_PN_fluxbalance} have been explicitly checked \emph{per se}, but also retrieved from the expression of the acceleration.\footnote{For instance, one is able to reconstruct the Hamiltonian flux-balance equation from the expression of $m_1v_1^ia_1^i+m_2v_2^ia_2^i$.}
Explicit expressions for $\mathcal{P}^i$ and $\mathcal{G}^i$ are given in~\cref{eq_lengthy_Pi,eq_lengthy_Gi} (the expressions for $\mathcal{H}$, $\mathcal{J}^i$ and associated flux are given once reduced to the center-of-mass frame).

\subsection{Center-of-mass reduction}\label{subsec:CM}

To reduce our expressions in the center-of-mass (CoM) frame, we first have to properly define it, which is done by finding an appropriate coordinate change $(y_1^i,y_2^i) \to (x^i,X_\text{CoM}^i)$ in which the position of the CoM, $X_\text{CoM}^i$, can be set to vanish, and $x^i \equiv y_1^i-y_2^i$.
We will use the canonical combinations $m = m_1+m_2$, $\nu = m_1m_2/m^2$, $\delta = (m_1-m_2)/m$, $r = \vert x^i\vert$ and $n^i =x^i/r$, and keep the EM moments untouched for simplicity.
As the center-of-mass integral $\mathcal{G}^i$ is not conserved, one can not simply identify as done usually $m\,X_\text{CoM}^i = \mathcal{G}^i$, as setting $X_\text{CoM}^i = 0$ would violate the flux-balance equations~\eqref{eq_PN_fluxbalance}.
Instead, we have to set $m\,X_\text{CoM}^i = \mathcal{G}^i + \int\!\dd t \Psi^i$, whose cancellation is compatible with the flux-balance equations, together with the usual condition $\mathcal{P}^i = 0$.
This yields the coordinate change
\begin{equation}\label{eq_PN_y1_CoM}
y_1^i =  \frac{m_2}{m}\,x^i + z_\text{pp}^i + z_\text{EM}^i - \int\!\dd t\, \frac{\Psi^i}{m}\,,
\qquad\text{and}\qquad
y_2^i =  -\frac{m_1}{m}\,x^i + z_\text{pp}^i + z_\text{EM}^i - \int\!\dd t\, \frac{\Psi^i}{m}\,,
\end{equation}
where the point-particle sector, $z_\text{pp}^i$, is to be found \emph{e.g.} in Eq.~(B4) of~\cite{Bernard:2017ktp} and the EM sector is given at the required order by
\begin{subequations}
\begin{align}
&
z^i_\text{EM} = \frac{\alpha_\text{EM}\,G}{2m r^3\,c^4}\bigg[\delta\,\bigg( (q_1q_2)-3(nq_1)(nq_2)\bigg) x^i + (xq_2)\,q_1^i- (xq_1)\,q_2^i \bigg]\,,\\
&
\Psi^i =  \frac{\alpha_\text{EM}\,G}{r^3\,c^4}\Bigg[
\bigg(\frac{1+\delta}{2}(\dot{q}_1 q_2)+3\,\frac{1-\delta}{2}(n\dot{q}_1)(nq_2) + \frac{3\delta}{2r}(q_1 q_2)(nv) - \frac{3}{r} \,\dJ_1^{ab}n^aq_2^b  \bigg) x^i
- \frac{\delta}{2}(q_1q_2) v^i\nonumber\\
& \qquad\qquad\qquad
+\bigg((x\dot{q}_2)+\frac{1+\delta}{2}(v q_2)-3\,\frac{1+\delta}{2}(nq_2)(nv)\bigg) q_1^i
+ \dJ_1^{ij}q_2^j
+ \left(1 \leftrightarrow2\right) \Bigg]\,,
\label{eq_PN_CoM_Psi}
\end{align}
\end{subequations}
were parenthesis are shortcuts for scalar product, \emph{e.g.} $(q_1q_2) \equiv q_1^kq_2^k$.
Let us note that, as expected, the center-of-mass position can be located out of the $(x^i,v^i)$ plan, due to the EM force.

Although it may seem dangerous, the non-locality (in time) present in~\cref{eq_PN_y1_CoM} is harmless.
Indeed, our theory does not depend on $\boldsymbol{y}_1$ or $\boldsymbol{y}_2$ alone, but always on the relative separation $\boldsymbol{y}_1-\boldsymbol{y}_2$, which is purely local.
Therefore (and we have naturally explicitly checked it), all quantities of interest (Lagrangian, acceleration, Noetherian quantities and associated flux) are purely local when expressed in the CoM frame.
Note also that, as a check, we have verified that the quantities reduced in the CoM frame agree with those computed from the Lagrangian reduced in the CoM frame.
The expressions for $\mathcal{H}$, $\mathcal{J}^i$ and associated flux are displayed in App.~\ref{app_Noether}.

\section{Toy model for circular orbits and numerical applications}\label{sec:quasiCirc}

\subsection{Constant and aligned moments: quasi-circular orbits}\label{subsec:quasiCirc}

In this project, we are interested in the impact of the magnetic dipoles on the motion of the two companions. Since our computation does not constrain the dynamics of the EM dipoles, we constrain the electric and magnetic dipoles in the following way: we set $q_1^i=q_2^i=0$ and assume that the magnetic dipoles are constant $\dd \dJ_A^{ij}/\dd t=0$ and both perpendicular to $n^i$ and $v^i$ at a given initial time. By doing so, the motion is planar at all time since $\Upsilon^i$ vanishes, and thus $\mathcal{J}_i\propto (\boldsymbol{n}\times \boldsymbol{v})^i$ is conserved. This allows to define the dimensionless constant magnetic dipole directed along the normal to the orbital plane $\boldsymbol{\ell}$
\begin{equation}
\widetilde{\J}_A^i \equiv G m^2 \epsilon_{ijk}\dJ_A^{jk} = \widetilde{\J}_A \ell^i\,.
\end{equation}
In this configuration, the EM sector of the relative acceleration reads to NLO
\begin{equation}
\boldsymbol{a}^\text{EM} = \frac{3G^3m^3\aem\,\tilde{\mu}_1\tilde{\mu}_2}{c^4\,\nu\,r^4}\bigg\lbrace1 + \frac{1}{c^2}\bigg[ \big(2\nu-1\big)v^2+ \frac{5\nu}{2}(nv)^2-\bigg(\frac{8}{3}+2\nu\bigg)\frac{Gm}{r}\bigg]\bigg\rbrace\boldsymbol{n}\,.
\end{equation}

Now we further simplify the model by assuming a quasi-circular motion. In this case, the relative acceleration and all relevant quantities of our system can be expressed uniquely in terms of the orbital frequency $\omega$, $\boldsymbol{a}^\text{circ} = - r \omega^2 \boldsymbol{n}$, where at the NLO, after posing $\gamma=G m /(rc^2)$, we find
\begin{equation}\label{eq:omega2gamma}
\omega^2= \frac{G m}{r^3}\left\{1+(-3+\nu)\gamma +\left(6 + \frac{41}{4}\nu+\nu^2\right)\gamma^2-\frac{3\aem\widetilde{\J}_1\widetilde{\J}_2}{\nu}\gamma^2\left[ 1+\left(-\frac{8}{3}+3\nu\right)\gamma\right]  \right\}.
\end{equation}
Next, we introduce the gauge invariant PN variable
\begin{equation}
x =\left(\frac{G m \omega}{c^3} \right)^{2/3},
\end{equation}
and invert the relation~\eqref{eq:omega2gamma} to obtain the value of $\gamma$ in terms of $x$. To the NLO, this relation reads
\begin{equation}
\gamma = x\left\{ 1+\left(1-\frac{\nu}{3}\right)x+\left(1- \frac{65}{12}\nu\right)x^2 +\frac{\aem \widetilde{\J}_1\widetilde{\J}_2}{\nu}x^2\left[ 1+ \left(\frac{13}{3}+\frac{2\nu}{3}\right)x \right] \right\}.
\end{equation}
Finally, we express the conserved energy and angular momentum as functions of $x$
\begin{subequations}\label{eq_circ_EJ}
\begin{align}
\label{eq_circ_E}
E^\text{circ} &= -\frac{m \nu c^2 x}{2}\bigg\lbrace 1 + \left( -\frac{3}{4} -\frac{\nu}{12} \right)x +\left(-\frac{27}{8}+\frac{19}{8}\nu-\frac{\nu^2}{24}\right)x^2\\
& \qquad\qquad\qquad\nonumber
+\frac{2 \aem\widetilde{\J}_1\widetilde{\J}_2}{\nu}x^2\left[ 1+\left( \frac{25}{6}+\frac{5\nu}{6} \right)x \right]\bigg\rbrace\,,\\
\label{eq_circ_Ji}
\mathcal{J}_i^\text{circ} &= \frac{Gm^2\nu}{c\sqrt{x}}\ell^i\,\bigg\lbrace 1+\left( \frac{3}{2} +\frac{\nu}{6} \right)x + \left(\frac{27}{8}-\frac{19}{8}\nu+\frac{\nu^2}{24}\right)x^2\\
& \qquad\qquad\qquad\nonumber
 -\frac{2\aem\widetilde{\J}_1\widetilde{\J}_2}{\nu}x^2\left[ 1+\left( \frac{10}{3}+\frac{2\nu}{3} \right)x  \right]\bigg\rbrace\,.
\end{align}
\end{subequations}
Of course, we recover the well-known expressions for the circular energy and angular momentum to the 2PN order for point-particles.\\

In order to define proper orbits beyond our simple approximation, one should fix the dynamical evolution of the EM moments.
For instance, one can follow the spirit of~\cite{Bourgoin:2021yzf} and set the magnetic dipole to be directed along the spin of the stars, which seems to be the case in physically relevant scenarii. As the dynamics of the spins is fully fixed, this would allow to properly define the trajectories of the system. 
Fixing the dynamics of the EM moments could bear interesting and non-trivial phenomenological implications.
For instance, in the case where the magnetic dipoles are constant but not aligned, $\Upsilon^i$ is non-zero which means that $\mathcal{J}_i$ is no longer conserved and thus the motion is no longer planar.
In such configuration, the 1PN coupling between the EM and gravitational interactions could also lead to orbital decay, with strong imprints on the gravitational waveform.  
Fixing the dynamics of the EM dipoles is let for future work.

\subsection{Numerical applications}\label{subsec:AN}

In order to estimate the relative impact of the EM effects with respect to the usual gravitational ones, let us focus in more detail on the quasi-circular ansatz developed in the previous section.
To do so, we consider a system of two white dwarfs, with physical properties given in the Table I of~\cite{Bourgoin:2022ilm}, that we partially reproduce in Table~\ref{tab_num_param}.
As for the magnitude of the (dimensionfull) magnetic dipole, we follow~\cite{Pablo_2019} and take
\begin{equation}
\vert \mu_A^i \vert = \frac{2\pi}{\mu_0}\,B_A R_A^3\,,
\end{equation}
where $R_A$ is the equatorial radius of the star, and $B_A$ is the magnitude of the magnetic field at its surface.

\begin{table}[!h]
    \centering
    \begin{tabular}{|c|c|c|c|}
    \hline
   Parameter & Unit &  Value for the first star & Value for the second star \\
    \hline
    \hline
    $m_A$ & $M_\odot$ & $1.2$ & $0.3$ \\
    \hline
    $R_A$ & km & $6.0\cdot 10^3$ & $15 \cdot 10^3$ \\
    \hline
    $B_A$ & $G$ & $1.0\cdot10^9$ & $1.0\cdot10^9$ \\
    \hline
    \end{tabular}
    \caption{Numerical values taken for the physical parameters of each star, reproduced from~\cite{Bourgoin:2022ilm}. $m_A$ is the mass of the star $A$, $R_A$ its equatorial radius, and $B_A$ the magnitude of the magnetic field at its surface.}
    \label{tab_num_param}
\end{table}

One can first compare the leading EM effect to the purely 2PN contribution in the frequency~\eqref{eq:omega2gamma}, in a gauge-independent manner. With the numerical values of Table~\ref{tab_num_param}, we find
\begin{equation}\label{eq_circ_oemga2_num}
\vert \omega^2_\text{EM} \vert \simeq 9 \,\vert \omega^2_{2\text{PN}}\vert\,,
\end{equation}
indicating that the leading EM effect contributes more to the dynamics than the purely 2PN contribution.
Pushing our model further, we have considered three configurations with gravitational frequencies of respectively $10^{-1}$, $10^{-2}$ and $10^{-3}$ Hz (as we specified to circular orbits, those correspond to twice the orbital frequencies of the systems considered).
Those configurations have been chosen following~\cite{Bourgoin:2022ilm}, and probe the LISA frequency range.
For each of the configurations, we computed the magnitude of the purely EM effect in the energy~\eqref{eq_circ_E} and angular momentum~\eqref{eq_circ_Ji}, and compared it to the point-particle 1PN and 2PN coefficients.
The results, displayed in Table~\ref{tab_num_res} show that, for all configurations, the purely EM effects is weaker than the 1PN point-particle one, but stronger than the 2PN one, corroborating the numerical results for the frequency~\eqref{eq_circ_oemga2_num}.
This clearly indicates that taking into account the EM effect will be important for both the calibration of the LISA instrument,\footnote{Note that the numerical estimate~\eqref{eq_circ_oemga2_num} as well as the second and fourth lines of Table~\ref{tab_num_res} are independent of the frequency, and thus holds also for ET.} and the characterization of galactic binaries.
Note that those number are in agreement with the conclusions of~\cite{Bourgoin:2022ilm}, and reinforce the motivation for a proper inclusion of the dynamics of the EM moments.

\begin{table}[!h]
    \centering
    \begin{tabular}{|c|c|c|c|}
    \hline
     & Configuration 1 & Configuration 2 & Configuration 3 \\
    \hline
    \hline
    $\vert E_\text{EM}/E_{1\text{PN}}\vert$ & $3.2\cdot 10^{-3}$ & $6.8\cdot 10^{-4}$ & $1.5\cdot 10^{-4}$ \\
    \hline
    $\vert E_\text{EM}/E_{2\text{PN}}\vert$ & $16$ & $16$ & $16$ \\
    \hline
    \hline 
    $\vert \mathcal{J}^i_\text{EM}/\mathcal{J}^i_{1\text{PN}}\vert$ & $1.6\cdot 10^{-3}$ & $3.4\cdot 10^{-4}$ & $7.4\cdot 10^{-5}$ \\
    \hline
    $\vert \mathcal{J}^i_\text{EM}/\mathcal{J}^i_{2\text{PN}}\vert$ & $16$ & $16$ & $16$ \\
    \hline
    \end{tabular}
    \caption{Comparison between the EM effects and the usual PN corrections for point-particles, for the three configurations studied. Here, we contrast the EM to the purely 1PN and 2PN coefficients entering the energy and angular momentum~\eqref{eq_circ_EJ}.}
    \label{tab_num_res}
\end{table}

\section{Summary and conclusion}\label{sec:CCL}

The inclusion of environmental effects is of prime importance for the detection and characterization of sources that will be resolvable by future gravitational-wave detector, such as LISA or ET.
Among those effects, the intense magnetic fields born by white dwarfs play a major role, as double white dwarfs will be used for calibrating LISA.
But, to our knowledge, there existed no proper post-Newtonian treatment including electromagnetic effects beyond electric charges.
This work filled this gap by constructing an action describing the full electromagnetic structure of a celestial body.
The kinetic sector of this action is given by the usual gauged Einstein-Hilbert and Maxwell terms, resp.~\cref{eq_action_grav,eq_action_EM}, and the matter action is constructed in the spirit of effective theories: it consists of a point-particle ansatz~\cref{eq_action_pp}, dressed up with a set of electromagnetic moments~\cref{eq_action_matter_EM}.

Specifying it to the dipolar order by matching the effective matter action as~\cref{eq_action_mat}, we obtained a Fokker Lagrangian at the NNLO order, from which we derived the acceleration and Noetherian quantities, and reduced them in the center-of-mass frame.
However, as the dynamics of the EM moments is unconstrained, the Noeherian quantities are not conserved, but obey the flux-balance equations~\eqref{eq_PN_fluxbalance}.
In order to have an idea of the relative strength of the EM effect for realistic systems of double white dwarfs, we imposed a toy-model quasi-circular orbit with constant magnetic moments normal to the orbital plane, and used values displayed in Table~\ref{tab_num_param}.
In this configuration, the Noetherian quantities are conserved and the motion is planar.
We found that the magnetic correction to the frequency, energy and angular momentum are smaller than the 1PN effect for point particles, but larger than the 2PN one, see Table~\ref{tab_num_res}.
It may thus be important to take magnetic effects into account for the detection and characterization of sources by LISA.

This work focused on the Noetherian quantites, the natural next step would thus be to study the dissipative sector, and to derive the EM corrections to the gravitational phase and amplitude.
Nevertheless, the toy-model we used for the orbit is only valid in the restrictive case where the magnetic dipoles are normal to the orbital plane.
If the ansatz we used is fine enough to provide numerical estimates, extracting physical behavior naturally calls for a finer modeling of the orbit, allowing for any possible direction of the moments.
In such generic configuration, we can expect both orbital decay and precession happening at a relative 1PN order (as the leading order magnetic force is conservative).
This would require the knowledge of the dynamics for the EM moments, which can be done by including the spins of the two stars.
This implementation, that needs to be achieved before turning to the dissipative sector, is left for future work.

\acknowledgments

It is a pleasure to thank M.-C. Angonin, G. Faye, M. Khalil and J. Steinhoff for enlightening discussions.
F.L. received funding from the European Research Council (ERC) under the European Union’s Horizon 2020 research and innovation program (grant agreement No 817791). 
C.LPL. acknowledges the financial support of Centre National d'\'{E}tudes Spatiales (CNES) for LISA.

\appendix

\section{Lengthy expressions}\label{app:LenExpr}

\subsection{Potential definitions}\label{app:PNpots}

The explicit expressions entering the wave equations~\eqref{eq:potdefs} read to consistent order
\begin{subequations}\label{eq:potsourcesdef}
\begin{align}
\xi_A &= m_A\left[ 1+\frac{3v_A^2-2V_A}{2c^2} +\frac{\varphi_A^2-2\left(W_{kk}^\text{EM}\right)_A}{\aem c^6} \right]-\frac{\dJ_A^{ab} (\partial_a\chi_b)_A}{c^6} \nn \\
&\quad+\frac{q_A^a}{c^4}\left[ -(\partial_a\varphi)_A +\frac{1}{c^2}\left( (\partial_a\varphi)_A\left(2V_A-\frac{3}{2}v_A^2  \right)+(\partial_t\chi_a)_A + (\partial_a V)_A\varphi_A \right)\right]\,,\\
\overline{\xi}_A^i &=\frac{q_A^i\varphi}{c^4}\left( 1-\frac{(4V +2V_A+v_A^2)}{2c^2} \right) \,,\\
\zeta_A &= m_A -\frac{q_A^a(\partial_a \varphi)_A }{c^4} \,,\\
\lambda_A &= m_A \,,\\
\omega_A &= \frac{m_A\varphi_A}{\aem c^2}\left(1 +\frac{3v_A^2-4V_A}{2c^2} \right) -\frac{4V_A^a\dot{q}_A^a}{c^4}\,,\\
\overline{\omega}_A^i &= q_A^i\left[ 1 -\frac{2V+2V_A+v_A^2}{2c^2}+ \frac{v_A^2(4(V-3V_A)-v_A^2)+4(2V V_A+V_A^2+8v_A^aV_A^a-4W_{kk})}{8c^4} \right]\nn\\
&\quad + \frac{4}{c^4}\left(v_A^i q_A^aV^a + \dJ^{ia}_AV^a\right)\,,\\
\theta_A^i &=\dot{q}_A^i\left[ 1+\frac{1}{c^2}\left(V_A- \frac{v_A^2}{2}\right) \right]
-\frac{q_A^i}{c^2}\partial_t\left(V_A +\frac{v_A^2}{2}\right)-\frac{4m_Av_A^i\varphi_A}{\aem c^2} \,,\\
\overline{\theta}_A^{ij} &=(\dJ_A^{ij}-q_A^iv_A^j)\left(1+\frac{4V-2V_A-v_A^2}{2c^2}  \right) \,,
\end{align}
\end{subequations}
where $W_{ij}^\text{EM}$ is defined by: $W_{ij} = W_{ij}^\text{pp} + W_{ij}^\text{EM}/(\aem c^4)$.

\subsection{Neotherian quantities}\label{app_Noether}

The complete expressions of the Noetherian quantities and associated fluxes at the required PN order are quite large, and not always very enlightening.
Therefore, we will only present them herebelow up to the $\calO(c^{-4})$ order, and let the interested reader refer to the ancillary file~\cite{AncFile} for their full-length expressions. We recall that in our notation, parenthesis denotes scalar product, \emph{e.g.} $(v_1v_2) \equiv v_1^kv_2^k$.\\

The quantity $Z^i$ entering in the boosted Lagrangian~\eqref{eq_PN_Noether_transfo_L} is given by $Z^i = Z_1^i + Z_2^i$ where
\begin{equation}\label{eq_lengthy_Zi}
Z_1^i = Z^i_{1,\text{pp}} + \frac{\alpha_\text{EM}}{c^4}\,Z^i_{1,\text{EM},\text{LO}}+ \frac{\alpha_\text{EM}}{c^6}\,Z^i_{1,\text{EM},\text{NLO}}\,,
\end{equation}
where
\begin{subequations}
\begin{align}
Z^i_{1,\text{pp}} = &\
- m_1y_1^i
+ \frac{m_1}{2\,c^2}\bigg(v_1^2+ \frac{Gm_2}{r_{12}}\bigg)y_1^i\\
&\ \nonumber
+ \frac{m_1}{c^4}\bigg[
\frac{v_1^4}{8}\,y_1^i + \frac{G m_2}{4\,r_{12}}\bigg( 2v_1^2-7(v_1v_2)+(v_1n_{12})^2-(v_1n_{12})(v_2n_{12}) \bigg)y_1^i\\
&\qquad\qquad \nonumber
+\frac{Gm_1}{r_{12}}\bigg(v_1^2- \frac{(v_1n_{12})^2}{4}\bigg)y_1^i- \frac{G^2m_1m_2}{2\,r_{12}^2}\,y_1^i- \frac{Gm_2(v_1n_{12})}{4}\,v_1^i\bigg]
+\left(1\leftrightarrow2\right)\,,\\
Z^i_{1,\text{EM},\text{LO}} = &\
\frac{G}{2\,r_{12}^3}\bigg[(q_2y_1)q_1^i-(q_1y_1)q_2^i-(q_1q_2)y_1^i+3(q_1n_{12})(q_2n_{12})y_1^i\bigg]\,,
\end{align}
\end{subequations}
and $Z^i_{1,\text{EM},\text{NLO}}$ is too long to be displayed here, but can be found in the ancillary file. The quantity $Z_2^i$ is naturally given by $Z_1^i$ under the exchange of the particles.\\

The linear momentum $\mathcal{P}^i$~\eqref{eq_PN_Noether_def_Pi} reads
\begin{equation}\label{eq_lengthy_Pi}
\mathcal{P}^i = \mathcal{P}^i_\text{pp} + \frac{\alpha_\text{EM}}{c^4}\,\mathcal{P}^i_{\text{EM},\text{LO}} + \frac{\alpha_\text{EM}}{c^6}\,\mathcal{P}^i_{\text{EM},\text{NLO}}\,,
\end{equation}
where the point-particle sector, $\mathcal{P}^i_\text{pp}$, is given \emph{e.g.} in Eq.~(4.3) of~\cite{deAndrade:2000gf},
\begin{equation}
\begin{aligned}
\mathcal{P}^i_{\text{EM},\text{LO}} =
\frac{G}{r_{12}^3} \bigg[ 
&
\frac{3}{2}\bigg( 
(n_{12}q_1)(v_1q_2) +(n_{12}q_2)(v_1q_1) - (n_{12}v_1)(q_1q_2)-5(n_{12}q_1)(n_{12}q_2)(n_{12}v_1)\\
& \qquad
+ \frac{r_{12}}{3} (\dot{q}_1 q_2) + (y_{12}\dot{q_1})(n_{12}q_2)
+2 \dJ_1^{ab}q_2^an_{12}^b
\bigg) n_{12}^i
 \\
&
- \frac{3}{2}\bigg((q_1q_2)-(n_{12}q_1)(n_{12}q_2) \bigg)v_1^i
- \frac{1}{2}(y_{12}q_2)\dot{q}_1^i
+ \dJ_1^{ia}q_2^a\\
&
+ \bigg( (y_{12}\dot{q}_2) +\big(3(n_{12}q_2)n_{12}^a -q_2^a\big)\big(v_1^a+v_2^a\big)\bigg)\frac{q_1^i}{2}
\bigg] + \left(1\leftrightarrow2\right)\,,
\end{aligned}
\end{equation}
and $\mathcal{P}^i_{\text{EM},\text{NLO}}$ is displayed in the ancillary file. We recall that no flux is associated with the linear momentum.\\

Similarly, the center-of-mass integral $\mathcal{G}^i$~\eqref{eq_PN_Noether_def_Pi} reads
\begin{equation}\label{eq_lengthy_Gi}
\mathcal{G}^i = \mathcal{G}^i_\text{pp} + \frac{\alpha_\text{EM}}{c^4}\,\mathcal{G}^i_{\text{EM},\text{LO}} + \frac{\alpha_\text{EM}}{c^6}\,\mathcal{G}^i_{\text{EM},\text{NLO}}\,,
\end{equation}
where the point-particle sector, $\mathcal{G}^i_\text{pp}$, is given \emph{e.g.} in Eq.~(4.5) of~\cite{deAndrade:2000gf},
\begin{equation}
\mathcal{G}^i_{\text{EM},\text{LO}} =
\frac{G}{2r_{12}^3} \bigg[ 
(q_1q_2)y_1^i - 3(n_{12}q_1)(n_{12}q_2)y_1^i-(y_{12}q_2)q_1^i
\bigg] + \left(1\leftrightarrow2\right)\,,
\end{equation}
and $\mathcal{G}^i_{\text{EM},\text{NLO}}$ is to be found in the ancillary file. The associated flux, $\Psi^i$ is displayed at leading order in~\cref{eq_PN_CoM_Psi}, and its NLO is stored in the ancilliary file.\\

In the center-of-mass frame and a 2PN relative, the Hamiltonian~\eqref{eq_PN_Noether_def_H} is given by
\begin{equation}\label{eq_lengthy_H}
\mathcal{H} = \mathcal{H}_\text{pp} + \frac{\alpha_\text{EM}}{c^2}\,\mathcal{H}_{\text{EM},\text{LO}} + \frac{\alpha_\text{EM}}{c^4}\,\mathcal{H}_{\text{EM},\text{NLO}} + \frac{\alpha_\text{EM}}{c^6}\,\mathcal{H}_{\text{EM},\text{NNLO}}\,,
\end{equation}
where $\mathcal{H}_\text{pp}$ is given \emph{e.g.} in Eq.~(3.1) of~\cite{Bernard:2017ktp},
\begin{subequations}
\begin{align}
\mathcal{H}_{\text{EM},\text{LO}} = & 
-\frac{G}{r^3}\bigg[3(nq_1)(nq_2)- (q_1q_2)\bigg]\,,\\
\mathcal{H}_{\text{EM},\text{NLO}} = 
& \nonumber
-\frac{G}{4r^3}\bigg[(3\nu-1)v^2(q_1q_2)+ 3(1-\nu)v^2(nq_1)(nq_2)-2\nu(vq_1)(vq_2)\\
& \qquad\qquad\nonumber
-3\nu (q_1q_2)(nv)^2 - 15\nu(nq_1)(nq_2)(nv)^2
+ 12\nu (nq_1)(vq_2)(nv)\\
& \qquad\qquad\nonumber
+(x\dot{q}_1)(x\dot{q}_2)+r^2(\dot{q}_1\dot{q}_2)
+2 \dJ_1^{ab}\dJ_2^{ab} + 6\dJ_1^{ab}\dJ_2^{ac}n^bn^c -4 \dot{\dJ}_1^{ab}q_2^ax^b
\bigg]\\
&
+ \frac{G^2 m}{2r^4}\bigg[7(nq_1)(nq_2)- 2(q_1q_2)+\frac{1-\delta}{2}\,(nq_1)^2\bigg]
+\left(1\leftrightarrow2\right)\,,
\end{align}
\end{subequations}
and $\mathcal{H}_{\text{EM},\text{NNLO}}$ is to be found in the ancillary file.
The associated flux reads
\begin{equation}\label{eq_lengthy_F}
\mathcal{F} = \frac{\alpha_\text{EM}}{c^2}\,\mathcal{F}_{\text{LO}} +\frac{\alpha_\text{EM}}{c^4}\,\mathcal{F}_{\text{NLO}} + \frac{\alpha_\text{EM}}{c^6}\,\mathcal{F}_{\text{NNLO}}\,,
\end{equation}
with
\begin{subequations}
\begin{align}
\mathcal{F}_{\text{LO}} = & 
\frac{G}{r^3}\big[3 n_{ab}-\delta_{ab}\big] \frac{\dd(q_1^aq_2^b)}{\dd t}\,,\\
\mathcal{F}_{\text{NLO}} = 
& \nonumber
\frac{G}{4r^3}\bigg[
\bigg(2\nu v_{ab}+12\nu(nv)n_av_b-3(1-\nu)v^2n_{ab}+(1-3\nu)v^2\delta_{ab}\\
& \nonumber \qquad \qquad
+ 3\nu (nv)^2\big(5n_{ab}+\delta_{ab}\big) \bigg)\frac{\dd(q_1^a q_2^b)}{\dd t}
+ \bigg(x_{ab}-r^2\delta_{ab}\bigg)\frac{\dd(\dot{q}_1^a\dot{q}_2^b)}{\dd t}\\
& \nonumber  \qquad
+ (1+\delta)\bigg(x_av_b+x_bv_a-3(xv)n_{ab}-(xv)\delta_{ab} \bigg)\frac{\dd(\dot{q}_1^a q_2^b)}{\dd t}\\
& \nonumber \qquad
+3(1-\delta)\bigg(3(nv)n_a-v_a\bigg)\frac{\dd(\dJ_1^{ak} q_2^{k})}{\dd t}
+4 x_a\frac{\dd(\dot{\dJ}_1^{ak} q_2^k)}{\dd t}
- 2\bigg(3n_{ab}-\delta_{ab}  \bigg)\frac{\dd(\dJ_1^{ak} \dJ_2^{bk})}{\dd t}\bigg]\\
& 
- \frac{G^2m}{4r^3}\bigg[(1-\delta)n_{ab}\frac{\dd(q_1^aq_1^b)}{\dd t}+2(7n_{ab}-2\delta_{ab})\frac{\dd(q_1^a q_2^b)}{\dd t}\bigg]
+\left(1\leftrightarrow2\right)\,,
\end{align}
\end{subequations}
and $\mathcal{F}_{\text{NNLO}}$ is stored in the ancillary file.\\

Lastly, the angular momentum is given in the center-of-mass frame by
\begin{equation}\label{eq_lengthy_Ji}
\mathcal{J}^i = \mathcal{J}^i_\text{pp} + \frac{\alpha_\text{EM}}{c^4}\,\mathcal{J}^i_{\text{EM},\text{LO}} + \frac{\alpha_\text{EM}}{c^6}\,\mathcal{J}^i_{\text{EM},\text{NLO}}\,,
\end{equation}
where $\mathcal{J}^i_\text{pp}$ is given \emph{e.g.} in Eq.~(3.2) of~\cite{Bernard:2017ktp},
\begin{equation}
\begin{aligned}
\mathcal{J}^i_{\text{EM},\text{LO}} =
\frac{G}{r^2}\bigg[
& 
\bigg( \frac{1-3\nu}{2}(q_1q_2) -\frac{3(1-\nu)}{2}(nq_1)(nq_2)\bigg) \varepsilon_{ijk}n^jv^k\\
& +\bigg(\nu (vq_2)- 3\nu (nq_2)(nv)+ \frac{1-\delta}{4}(x\dot{q}_2)\bigg) \varepsilon_{ijk}n^jq_1^k\\
&
+ \frac{1+\delta}{4}(xq_2)\,\varepsilon_{ijk}n^j\dot{q}_1^k
+ \frac{1-\delta}{2}\,\varepsilon_{ijk}n^j\J_1^{ka}q_2^a\bigg] +\left(1\leftrightarrow 2\right)\,,
\end{aligned}
\end{equation}
and $\mathcal{J}^i_{\text{EM},\text{NLO}}$ is to be found in the ancillary file.
The associated flux reads
\begin{equation}\label{eq_lengthy_Upsiloni}
\Upsilon^i =\frac{\alpha_\text{EM}}{c^2}\,\Upsilon^i_{\text{LO}} +\frac{\alpha_\text{EM}}{c^4}\,\Upsilon^i_{\text{NLO}} + \frac{\alpha_\text{EM}}{c^6}\,\Upsilon^i_{\text{NNLO}}\,,
\end{equation}
with
\begin{subequations}
\begin{align}
\Upsilon^i_{\text{LO}} =
& 
- \frac{3G}{r^3}\,(nq_2)\,\varepsilon_{ijk}n^jq_1^k + \left(1 \leftrightarrow 2\right)\,,\\
\Upsilon^i_{\text{NLO}} =
&  \nonumber
\frac{G}{r^3}\bigg[ \frac{3}{2}\bigg((1-\nu)v^2(nq_2)-5\nu(nq_2)(nv)^2+2\nu(vq_2)(nv)\\
& \qquad\quad+\frac{1-\delta}{6}\big(3(x\dot{q}_2)(nv)-r(v\dot{q}_2)\big)\bigg)\varepsilon_{ijk}n^jq_1^k\\
& \qquad \nonumber
+ \bigg(3\nu(nq_2)(nv)-\nu(vq_2) - \frac{1-\delta}{4}(x\dot{q}_2)\bigg)\varepsilon_{ijk}v^jq_1^k
- \frac{1+\delta}{4}(xq_2)\,\varepsilon_{ijk}v^j\dot{q}_1^k\\
& \qquad \nonumber
+ \frac{r}{2}\bigg(\frac{1+\delta}{2}\big(3(nq_2)(nv)-(vq_2)\big) -(x\dot{q}_2)\bigg)\varepsilon_{ijk}n^j\dot{q}_1^k\\
& \qquad \nonumber
+3\bigg(\frac{1-\delta}{2}(nv)q_2^a-n_b\dJ_2^{ab}\bigg)\varepsilon_{ijk}n^j\dJ_1^{ka}
- r\, q_2^a \,\varepsilon_{ijk}n^j\dot{\dJ}_1^{ka}
- \frac{1-\delta}{2}\, q_2^a \,\varepsilon_{ijk}v^j\dJ_1^{ka}\bigg]
\\
&
+ \frac{G^2m}{2r^4}\bigg[14(nq_2) + (1-\delta)(nq_1)\bigg] \,\varepsilon_{ijk}n^jq_1^k + \left(1 \leftrightarrow 2\right)\,,
\end{align}
\end{subequations}
and $\Upsilon^i_{\text{NNLO}}$ is stored in the ancillary file.

\bibliography{ListeRef_GWmag.bib}

\end{document}